\documentclass[aps,pra,twocolumn,showpacs,preprintnumbers,amsmath,amssymb]{revtex4-2}
\usepackage{soul}
\usepackage{amsmath}
\usepackage{esvect}

\usepackage{natbib}
\usepackage{graphicx}
\usepackage{color}
\usepackage{tabularx}
\usepackage{multirow}

\usepackage{wasysym}

\usepackage{amssymb}

\usepackage[utf8]{inputenc}

\usepackage{float}

\usepackage{subfigure}

\usepackage{float}

\usepackage{hyperref}

\usepackage{cancel}

\usepackage{xcolor}

\usepackage{setspace}
\usepackage{soul}
\usepackage{wrapfig}

\definecolor{burntorange}{rgb}{0.8, 0.33, 0.0}

\begin{document}

\title{Nuclear-spin-related properties of~the dual-frequency Doppler-free resonance}

\author{E.\,A.~Tsygankov$^{1}$}
\email[]{tsygankov.e.a@yandex.ru}
\author{K.\,M.~Sabakar$^{1}$}
\author{D.\,S.~Chuchelov$^{1}$}
\author{M.\,I.~Vaskovskaya$^{1}$}
\author{V.\,V.~Vassiliev$^{1}$}
\author{S.\,A.~Zibrov$^{1}$}
\author{V.\,L.~Velichansky$^{1}$}
\affiliation{1. P.\,N. Lebedev Physical Institute of the Russian Academy of Sciences,\\
Leninsky Prospect 53, Moscow, 119991 Russia}

\newcommand{\red}{\textcolor{red}}
\newcommand{\blue}{\textcolor{blue}}

\begin{abstract}
We~investigate the dual-frequency Doppler-free resonance in~the D$_1$ line of~alkali-metal atoms for any accessible value of~the nuclear spin $I$. The~consideration is~performed using the symmetries of~the dipole operator and the basis, where the quantization axis is~directed along the polarization of~the one of~optical waves. We~show that there is~the absence of~the optical pumping in~the scheme with parallel polarizations for~the center of~the crossover, resulting in~its smallest width. Secondly, the growth in~the absorption for the center of~the peak with \hbox{$F_e=I-1/2$} and the decrease of~its width with the two-photon detuning in~the case of~orthogonal polarizations is~explained. Particular attention is~paid to~the special case of~$I=3/2$, where this effect is~the most pronounced. \hbox{The experiment} with $^{87}$Rb, $^{85}$Rb, and $^{133}$Cs atoms is~in~agreement with the analysis.
\end{abstract}

\maketitle

\section{Introduction}

Doppler-free spectroscopy is~a~powerful tool providing narrow resonance, one application of~which is~the frequency stabilization of~the laser radiation~\cite{letokhov1977nonlinear,schawlow1982spectroscopy,hansch1977high}. Recently, a~compact optical frequency standard has been proposed using microresonator-based frequency combs to~transfer the frequency stability in~the consumer range~\cite{newman2019architecture}. As~a~reference resonance for such a~standard, the two-photon transition in~$^{87}$Rb atoms is~currently mainly used~\cite{PhysRevApplied.9.014019, PhysRevApplied.12.054063,Beard:24,10722354,Duspayev_2024}. However, the high-contrast dual-frequency Doppler-free resonance initially observed in~$^{133}$Cs~\cite{hafiz2016doppler,hafiz2017high,zhao2021laser,gusching2021short} also shows a~strong potential for this application. The demonstrated frequency stability of~the laser stabilized to~such resonance reached the level \hbox{of~$3\cdot10^{-13}$} at~1~s~\cite{gusching2023short}.

The term dual-frequency refers to~the use of~the bichromatic counter-propagating optical waves to~induce the resonance. The fields are generated through frequency-modulated laser radiation. 
Usually, when the modulation frequency is~set to~the half of~the ground-state hyperfine splitting, the first-order sidebands of~the spectrum are tuned to~the absorption line.
In~the case of~orthogonal linear polarizations of~the fields, inverted eigen peaks (associated with atomic transitions) with high amplitudes and narrow widths were observed in~$^{133}$Cs. These results were explained by~optical pumping processes associated with the coherent population trapping and Hanle effects~\cite{hafiz2016doppler}.

In our recent work~\cite{DFDF87Rb}, we~performed dual-frequency Doppler-free spectroscopy of~the $^{87}$Rb D$_1$ line and investigated the properties of~the resulting spectra in~the schemes with orthogonal (lin$\,\perp\,$lin) and parallel (lin$\,\parallel\,$lin) polarizations. In~the first case, we~observed high-contrast Doppler-free absorption eigen peaks, while parallel polarizations gave a~pronounced inverted crossover. Its width was the narrowest among the observed peaks. We~also explored the resonance characteristics as~a~function of~the two-photon detuning and have found the increase in~the amplitude of~the low-frequency eigen peak.

In~this paper, we~extend the previously reported results to~atoms with various nuclear spin values $I$, both integer (fermions) and half-integer (bosons). In~general case, we~analyze the behavior of~eigen peaks as~a~function of~the two-photon detuning and provide an~explanation for the narrowest linewidth of~the crossover resonance. The results of~the analysis are consistent with our experiment performed with $^{87}$Rb, $^{85}$Rb, and $^{133}$Cs atoms, which are the most frequently used and accessible. The consideration is~potentially applicable for Doppler-free spectroscopy of~$^{39,\,40,\,41}$K, $^{23}$Na, $^{7}$Li, and ions of~$^{171}$Yb.

\section{Theory}

This section presents a~sequential analysis of~the dual-frequency Doppler-free resonance related to~the transition \hbox{$J_g=1/2\rightarrow J_e=1/2$}. Crossover and eigen peaks are investigated in~\hbox{lin$\,||\,$lin} and \hbox{lin$\,\perp\,$lin} schemes, respectively. Their characteristics are considered for any accessible nuclear spin $I$.

\begin{figure*}[t] 
\centering
\includegraphics[width=\textwidth]{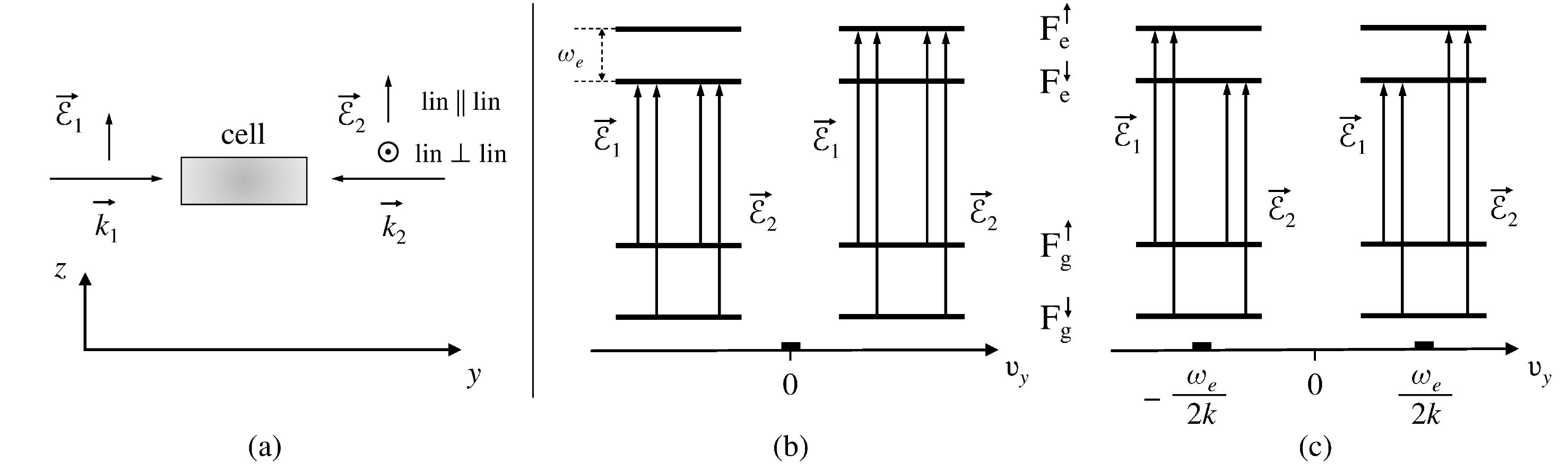}
  \caption{Optical fields diagram~(a). The energy level structure and transitions involved in~the formation of~the both eigen~(b) and crossover(c) 
  peaks. Below the level schemes, the corresponding atomic groups in~the longitudinal velocity space~$\upsilon_y$ contributing to~each peak are indicated. For eigen peaks, these groups are located in~the vicinity of~$\upsilon_y=0$. Two groups of~atoms with~$\upsilon_y=\pm\omega_e/2k$ contribute to~the crossover formation simultaneously. Here, $\omega_e$ and $k$ are the excited-state hyperfine splitting and the wave vector, respectively.}
  \label{Resonances_scheme}
\end{figure*}

The analysis of~all resonance features is~performed in~a~basis where the quantization axis is~aligned with the polarization vector of~one of~the optical fields $\vec{\mathcal{E}}_1$; see Fig.~\ref{Resonances_scheme}a. We~note that in~the more frequently used basis the quantization axis is~directed orthogonally to~polarizations of~the optical waves. In~this case, they both induce only $\sigma$ transitions and Zeeman coherences should be~accounted in~the analysis. The qualitative explanation of~the described below effects will not be~so~straightforward, if~it~is even possible. In~contrast, in~our basis, the field $\vec{\mathcal{E}}_1$ induces only $\pi$ transitions, therefore the Zeeman $\Lambda$-schemes are absent.

It~is~assumed that the electric field components of~the counter-propagating waves, $\vec{\mathcal{E}}_1$ and $\vec{\mathcal{E}}_2$, are equal in~amplitudes and oscillate in~phase, while propagation effects along the optical axis are neglected. In~this case, the absorption of~the optical fields is~equal due to~the symmetry of~the consideration. Therefore, we~analyze only the absorption of~the field $\vec{\mathcal{E}}_1$, which is~simpler than for $\vec{\mathcal{E}}_2$---the first field~does not induce Zeeman coherences, and we~do~not account for them in~the analysis.

Further, in~the text, we~use the following notation: subscripts $_g$, $_e$ denote the ground and excited states, respectively, and superscripts denote the upper ($^\uparrow$) or~lower ($^\downarrow$) hyperfine sublevels. In~the general case, $F^\uparrow\equiv I+1/2$, $F^\downarrow\equiv I-1/2$.

We~briefly outline the formation mechanisms of~the eigen and crossover peaks induced by~counter-propagating bichromatic optical fields.
Eigen peaks arise when the both waves interact with the only one group of~atoms with zeroth longitudinal velocity $v_y=0$. In~this case, the both fields induce transitions from the two ground-state levels to~a~common excited-state level; see Fig.~\ref{Resonances_scheme}b.
The crossover occurs when the waves interact with two velocity groups of~atoms with $v_y=\pm\omega_e/2k$, where $\omega_e$ is~the excited-state hyperfine splitting and $k$ is~the wave vector. Hence, both optical fields induce transitions to~each of~the excited-state levels, but act on~different velocity groups of~atoms; see Fig.~\ref{Resonances_scheme}c. 
The effective velocity range of atoms interacting with the optical fields is~given by~$\Delta v_y \simeq (\gamma / k)\sqrt{(1 + s)}$, where $\gamma$ is~the natural width, $s$ is~the saturation parameter.

The analysis is~structured as~follows. We~examine the interaction of~atoms with the optical fields out of~the optical resonance (in~the Doppler background) and at~the exact optical resonance (in~the line's center), firstly for the crossover. Secondly, we~consider this interaction for the low-frequency eigen peak both at~two-photon resonance and with the two-photon detuning, and lastly for the high-frequency eigen peak.

\subsection{The crossover}

\begin{figure}[t]
  \center
\includegraphics[width=1\columnwidth]{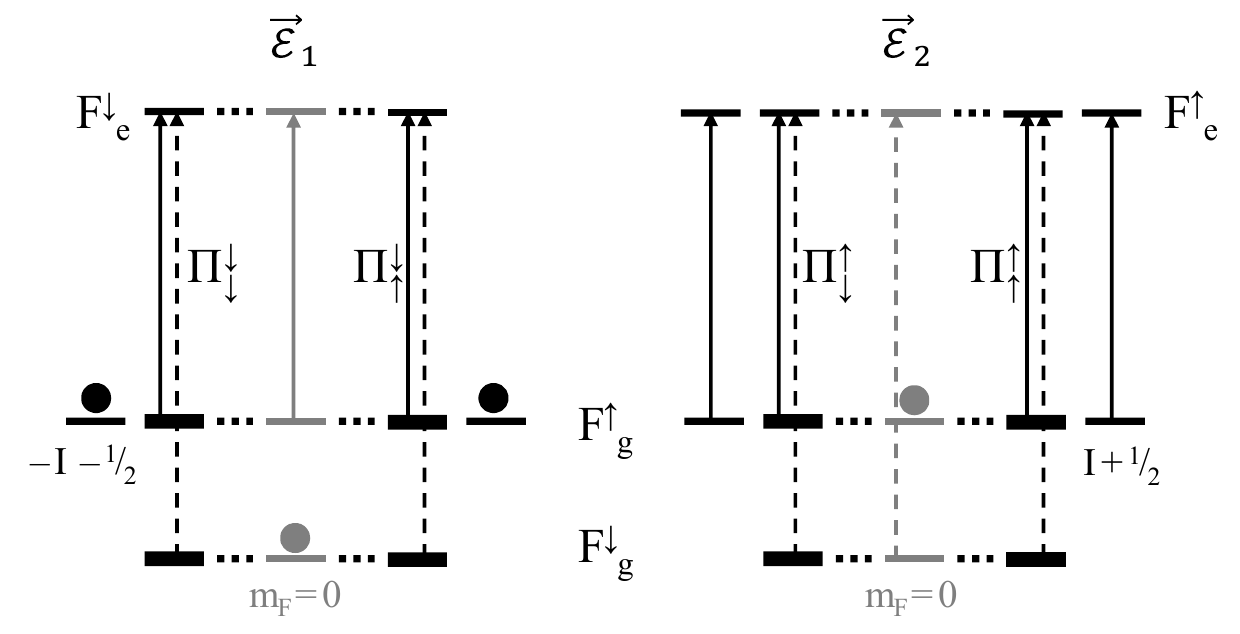}
   \caption{Schemes of~energy levels and $\pi$-transitions to~the $F^\downarrow_e$ (left) and $F^\uparrow_e$ (right) states induced by~the optical fields in lin$\,||\,$lin configuration for the case of~arbitrary value of~the nuclear spin~$I$. Vertical solid and dashed lines indicate transitions from $F^\uparrow_g$ and $F^\downarrow_g$, respectively. Bold lines show~the magnetic sublevels which are involved in~some of~the dark superpositions created by~each wave, horizontal dotted lines stand for other possible sublevels. Sublevels with $m_F = 0$, which exist only for half-integer~$I$, are shown in~gray. Circles denote non-absorbing sublevels that accumulate atomic population.}
  \label{Crossoverscheme}
\end{figure}

We begin our consideration for the crossover in~the lin$\,||\,$lin scheme. 
In~the chosen representation, there are only $\pi$ induced electric-dipole transitions that conserve the magnetic quantum number $m_F$. 
In~the non-resonant case, for the optical field interacting with atoms through the lower excited level $F^\downarrow_e$, there are always two non-absorbing sublevels $F^\uparrow_g, |m_{F_g}|=I+1/2$. 
For half-integer values of~$I$ (boson atoms), there is~a~third non-absorbing sublevel $F^\downarrow_g, m_{F_g}=0$. For the second wave interacting with the opposite velocity group of~atoms through the upper excited level $F^\uparrow_e$, there is~one non-absorbing sublevel $F^\uparrow_g, \,m_{F_g}=0$; see Fig.~\ref{Crossoverscheme}. 
Both optical waves also form non-absorbing superpositions of~states at~sublevels with the same $m_{F_g}$ of~different $F_g$, due to~the effect of~coherent population trapping.
Thus, absorption at~the wings of Doppler-broadened line is~suppressed due to~the optical pumping of~atoms into non-absorbing sublevels and dark superpositions.

At the optical resonance, both waves interact with the same velocity group of~atoms, so~all the sublevels become absorbing. It~can be~shown that, independently on~the $I$ value, the products of~the Rabi frequencies for $\Lambda$-schemes formed through different excited-state levels are of~the same absolute value but are opposite in~sign. Therefore, dark superpositions of~states are not formed. As~a~result, absorption increases markedly, which explains the inversion of~the crossover and its greater amplitude compared to~the lin$\,\perp\,$lin scheme. In~the case of~orthogonal polarizations, the field $\vec{\mathcal{E}}_2$ induces $\sigma$ transitions, therefore the non-absorbing sublevels and dark superpositions are present.

The combination of~large amplitude and narrow linewidth arises from the equal depopulation rate of~the ground-state sublevels and their isotropic repopulation by~the spontaneous emission. As~a~consequence, no~optical pumping occurs---the equilibrium distribution of~the population among ground-state sublevels is~conserved. To~demonstrate that this absence of~optical pumping does not depend on~the nuclear spin value, we~consider the reduction of~the dipole operator, which reads as:
\begin{equation}
\begin{gathered} 
\langle F_e,\,m_{F_e}|er_q|F_g,\,m_{F_g}\rangle\\
=\left[-1\right]^{F_g-1+m_{F_e}}\sqrt{2F_e+1}\begin{pmatrix}
F_g & 1 & F_e \\
m_{F_g} & q & -m_{F_e}
\end{pmatrix}\langle F_e\parallel e\mathbf{r}\parallel F_g\rangle\\
=[-1]^{2F_g+m_{F_e}+J_e+I}\sqrt{(2F_g+1)(2F_e+1)(2J_e+1)}\\
\cdot\begin{pmatrix}
F_g & 1 & F_e \\
m_{F_g} & q & -m_{F_e}
\end{pmatrix}\begin{Bmatrix}
J_e & J_g & 1 \\
F_g & F_e & I
\end{Bmatrix}\langle J_e\parallel e\mathbf{r}\parallel J_g\rangle\\
\equiv \mathcal{K}^{F_e,\,m_{F_e}}_{F_g,\,m_{F_g}}\langle J_e\parallel e\mathbf{r}\parallel J_g\rangle,
\end{gathered}
\label{DOReduction}
\end{equation}

\noindent where $\langle J_e\parallel e\mathbf{r}\parallel J_g\rangle$ is~the reduced dipole matrix element, the~parentheses are used to~denote the $3$-j symbol, and the braces---the $6$-j symbol. The coefficient $\mathcal{K}^{F_e,\,m_{F_e}}_{F_g,\,m_{F_g}}$ is~introduced for brevity.  In~our case, the $q$ component of~$\mathbf{r}$ in~the spherical basis is~equal to~zero, since the $\pi$ transitions are considered. The equality \hbox{$J_e=J_g=1/2$} also simplifies the consideration.

Let us~refer to~the induced electric-dipole transitions between the sublevels \hbox{$F^\downarrow_g,\,m_F$} and \hbox{$F^\downarrow_e,\,m_F$}.
In~this case, the first row of~the $3$-j symbol is~given by~\hbox{$\left(I - 1/2,\,1,\,I - 1/2\right)$}, and the second row by~\hbox{$\left(m_F,\,0,\,m_F\right)$}.
For the corresponding $6$-j symbol, the rows are $\left(1/2,\,1/2,\,1\right)$ and $\left(I - 1/2,\,I - 1/2,\,I\right)$.
Using the explicit form of~the $3$-j symbol and the Racah formulae for the $6$-j symbol, we~transform the term in~the front of the reduced dipole matrix element in~Eq.~\eqref{DOReduction} into
\begin{equation}
\begin{gathered}
\mathcal{K}^{I-1/2,\,m_F}_{I-1/2,\,m_F}\equiv\Pi^{\downarrow}_{\downarrow}(m_F)=\left[-1\right]^{3I+m_F-1/2}\sqrt{8I^2}\\
\cdot\left(\left[-1\right]^{-I-m_F-1/2}m_F\sqrt{\dfrac{2}{I\left(2I+1\right)\left(2I-1\right)}}\right)\\
\cdot\left\{\dfrac{\left[-1\right]^{-2I}}{\sqrt{6}}\sqrt{\dfrac{2I-1}{2I\left(2I+1\right)}}\right\}\\
\equiv-m_F\dfrac{2}{\sqrt{3}}\dfrac{1}{2I+1}.
\label{minusminusvertical}
\end{gathered}
\end{equation}

Here, we~retain the parentheses and braces to~clearly indicate the $3$-j and $6$-j symbols. The notation $\Pi^{\downarrow}_{\downarrow}(m_F)$ is~introduced for brevity in~the subsequent expressions, where the arrows indicate the levels of~ground (subscript) and excited (superscript) states involved in~the transitions; see Fig.~\ref{Crossoverscheme}.

The induced electric-dipole transitions between sublevels \hbox{$F^\downarrow_g,\,m_F$} and \hbox{$F^\uparrow_e,\,m_F$} can be~treated in~the same way. This yields:

\begin{equation}
\mathcal{K}^{I+1/2,\,m_F}_{I-1/2,\,m_F}\equiv\Pi^{\uparrow}_{\downarrow}(m_F)=\dfrac{1}{\sqrt{3}}\dfrac{\sqrt{\left(2I+1\right)^2-4m^2_F}}{2I+1},
\label{minusplusvertical}
\end{equation}

\noindent where the total phase factor $\left[-1\right]^{2\left(I+m_F\right)+1}$ is~always positive as~the sum $I+m_F$ is~always a~half-integer.

From the Eqs.~\eqref{minusminusvertical},~\eqref{minusplusvertical} follows
\begin{equation}
\left[\Pi^{\downarrow}_{\downarrow}(m_F)\right]^2
+\left[\Pi^{\uparrow}_{\downarrow}(m_F)\right]^2
=\dfrac{1}{3},
\label{1/3forI-1/2}
\end{equation}

\noindent which means that the optical waves depopulate sublevels of~\hbox{$F^\downarrow_g$} with the same rate.

Further, we~consider induced electric-dipole transitions from sublevels of~\hbox{$F^\uparrow_g$}. The corresponding coefficient for $F^\uparrow_e$ is
\begin{equation}
\Pi^{\uparrow}_{\uparrow}(m_F)=m_F\dfrac{2}{\sqrt{3}}\dfrac{1}{2I+1}.
\label{plusplusvertical}
\end{equation}

The coefficient $\Pi^{\downarrow}_{\uparrow}(m_F)$ can be~obtained from 
$\Pi^{\uparrow}_{\downarrow}(m_F)$ due to~the symmetries of~$3$-j and $6$-j symbols. The value of~the $6$-j symbol remains unchanged, as~it~is~invariant under any permutation of~its columns. The $3$-j symbol, however, acquires a~phase factor of~$-1$ due to~the following. First, changing the signs of~$m_{F_g}$ and $m_{F_e}$ in~the second row introduces a~coefficient $[-1]^{F_g+F_e}=\left[-1\right]^{2I}$.
Next, swapping the first and third columns adds another coefficient $\left[-1\right]^{F_g+1+F_e}=\left[-1\right]^{2I+1}$. 
The total phase factor is~therefore $\left[-1\right]^{4I+1}=-1$. Taking into account the change in~sign of~$\left[-1\right]^{2F_g+m_{F_e}+J_e+I}$ standing in~the front of the $3$-j and $6$-j symbols [see Eq.~\eqref{DOReduction}], we~get
\begin{equation}
\Pi^{\downarrow}_{\uparrow}(m_F)=\Pi^{\uparrow}_{\downarrow}(m_F)=\dfrac{1}{\sqrt{3}}\dfrac{\sqrt{\left(2I+1\right)^2-4m^2_F}}{2I+1}.
\label{plusminusvertical}
\end{equation}

Thus, from the obtained expressions for the $\Pi$ coefficients, the relation follows:
\begin{equation}
\Pi^{\downarrow}_{\downarrow}(m_F)\cdot\Pi^{\downarrow}_{\uparrow}(m_F)+\Pi^{\uparrow}_{\downarrow}(m_F)\cdot\Pi^{\uparrow}_{\uparrow}(m_F)=0,
\end{equation}

\noindent which confirms the previously stated absence of dark-state superpositions at~the exact optical resonance. 

Another relation 
\begin{equation}
\left[\Pi^{\downarrow}_{\uparrow}(m_F)\right]^2+\left[\Pi^{\uparrow}_{\uparrow}(m_F)\right]^2=\dfrac13,
\label{1/3forI+1/2}
\end{equation}

\noindent together with Eq.~\eqref{1/3forI-1/2} ensures that the ground-state populations remain at~their equilibrium values $1/\left[4\left(I+1/2\right)\right]$. 
It~follows that the population is~uniformly redistributed among all ground-state sublevels at~the exact optical resonance. This completely suppresses optical pumping, thereby reducing the effective saturation parameter $s$ and resulting in~the narrowest width of~the crossover.

We~remind that symmetry given by~Eqs.~\eqref{1/3forI-1/2} and \eqref{1/3forI+1/2} reflects the fact that far-detuned linearly-polarized light, for which the hyperfine splittings can be~neglected, interacts only with one component of~the dipole operator.

\begin{figure}[b]
  \center
\includegraphics[width=1\columnwidth]{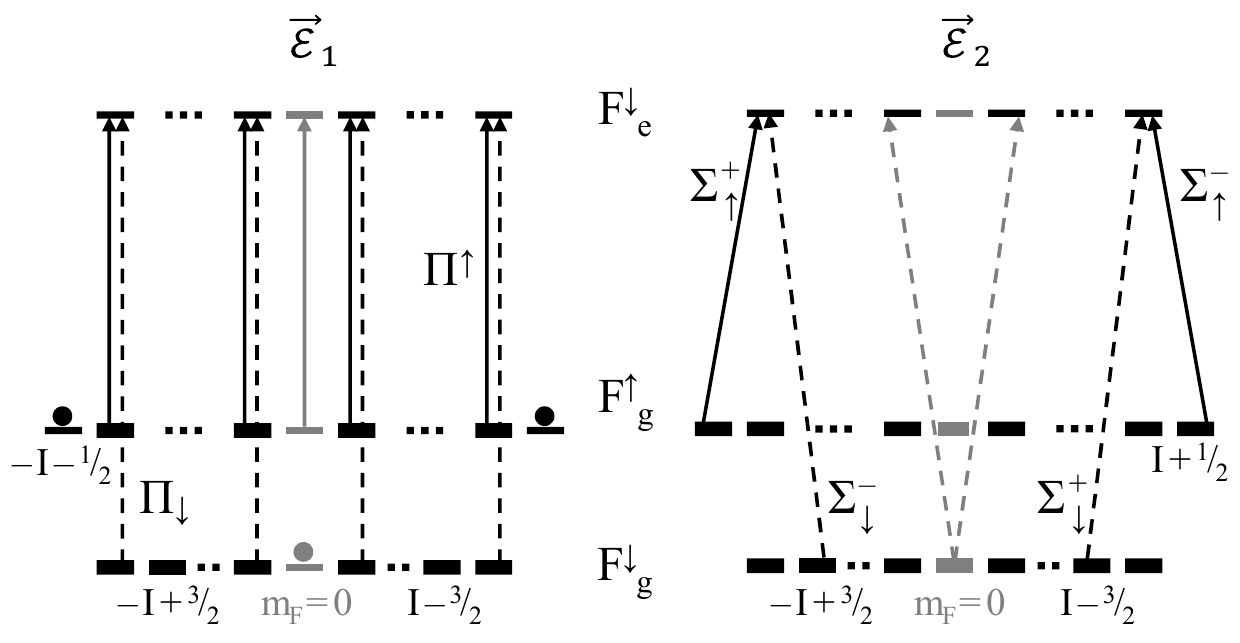}
  \caption{Schemes of~energy levels along with $\pi$ (left) and $\sigma$ (right) transitions to~$F^{\downarrow}_e$ state induced by~the optical fields in~the lin$\,\perp\,$lin configuration. For clarity, only some of~the allowed transitions are shown.}
  \label{Lowfreqscheme}
\end{figure}

\subsection{The low-frequency eigen peak}
\label{Low-Frequency-Section}

The peak formed by~transitions to~the $F^\downarrow_e$ level of~the excited state is~considered in~a~lin$\,\perp\,$lin configuration of~the optical fields. The field $\vec{\mathcal{E}}_1$ induces $\pi$-transitions, while field $\vec{\mathcal{E}}_2$ induces $\sigma$-transitions with~$\Delta m_F=\pm1$; see Fig.~\ref{Lowfreqscheme}.

In the Doppler background, the optical waves interacting with different atomic velocity groups pump atoms into non-absorbing sublevels and dark-state superpositions, analogous to~the case of~the crossover. A~clearer understanding can be~obtained by~focusing on~the absorption of~the wave $\vec{\mathcal{E}}_1$, which avoids the complications for the analysis introduced by Zeeman coherences in~the case of~$\vec{\mathcal{E}}_2$.

At the optical resonance, $\pi$ and $\sigma$ transitions are simultaneously induced in~the same velocity group of~atoms. To~analyze such interactions, in~addition to~the previously obtained coefficients $\mathcal{K}^{I-1/2,\,m_F}_{I-1/2,\,m_F}\equiv\Pi_{\downarrow}(m_F)$ and $\mathcal{K}^{I-1/2,\,m_F}_{I+1/2,\,m_F}\equiv\Pi_{\uparrow}(m_F)$ (we~omit the superscript, as~it~is~$\downarrow$ in~all cases throughout this subsection), one also needs the coefficients $\mathcal{K}^{I-1/2,\,m_F\pm1}_{I-1/2,\,m_F}\equiv\Sigma^\pm_{\downarrow}(m_F)$ and $\mathcal{K}^{I-1/2,\,m_F\pm1}_{I+1/2,\,m_F}\equiv\Sigma^\pm_{\uparrow}(m_F)$.
From their explicit form obtained through the procedure described in~the subsection above, it~follows that
\begin{equation}
\begin{gathered}
\bigg[\Pi_{\downarrow}\cdot\Pi_{\uparrow}\bigg](m_F)+\bigg[\Sigma^-_{\downarrow}\cdot\Sigma^-_{\uparrow}\bigg](m_F)+\bigg[\Sigma^+_{\downarrow}\cdot\Sigma^+_{\uparrow}\bigg](m_F)=0,
\label{PiAndSigmaProducts}
\end{gathered}
\end{equation}

\noindent i.e., the phases of dark-state hyperfine superpositions induced by~$\vec{\mathcal{E}}_1$ and $\vec{\mathcal{E}}_2$ are the opposite. As~a~result, the absorption of~$\vec{\mathcal{E}}_1$ from the sublevels with the same $m_F$ is~increased compared to~the Doppler background. Eq.~\eqref{PiAndSigmaProducts} is~physically meaningful only for $I>1/2$, since for $I = 1/2$ there are hyperfine levels \hbox{$F = 0,\,1$}, and no~hyperfine $\Lambda$-schemes can be~formed due to~the forbidden transition between $F_g = 0$ and $F_e = 0$. We~note that the last obtained symmetry reflects the fact that isotropic unpolarized light does not induce hyperfine coherences.

As can be~seen in~Fig.~\ref{Lowfreqscheme}, for $\vec{\mathcal{E}}_2$ exist hyperfine $\Lambda$-schemes formed at~sublevels that are non-absorbing for $\vec{\mathcal{E}}_1$. In~general case, two such schemes involve the transitions from ground sublevels \hbox{$F^\downarrow_g,\,m_F = |I - 3/2|$} and \hbox{$F^\uparrow_g,\,m_F = |I + 1/2|$} to~the excited sublevels \hbox{$F^\downarrow_e,\,m_F = |I - 1/2|$}; see Fig.~\ref{Lowfreqscheme}. Another two schemes arise only for boson atoms with half-integer $I$ and involve transitions from ground sublevels $F^\downarrow_g,\,m_F = 0$ and $F^\uparrow_g,\,m_F =\pm2$ to $F^\downarrow_e,\,m_F = \pm 1$. Fig.~\ref{Lowfreqscheme} shows only one transition for each scheme. The dark superpositions formed by these \hbox{$\Lambda$-schemes} trap atoms even at~the optical resonance. This effect is~the underlying reason for the dependence of~the considered eigen peak amplitude on~the two-photon detuning. We~call these~$\Lambda$-schemes affected, because one of~their ground-state levels ($m_F = |I - 3/2|$) is~connected by~$\pi$ transition induced by~the field~$\vec{\mathcal{E}}_1$ to~a~non-common excited-state level. Hence, even if~the relaxation in~the ground state is~absent, the common excited-state level becomes populated and the absorption is~increased.

In the Doppler background, the two-photon detuning destroys dark-state superpositions of~sublevels with $m_{F^\uparrow_g} = m_{F^\downarrow_g}$, resulting in~an~increased absorption. At~the optical resonance, the $\Lambda$-schemes responsible for trapping atoms at~the end magnetic sublevels (and on~the central one for bosons with the half-integer $I$) are also destroyed, and absorption increases. Depending on~the relative change in~absorption between resonant and off-resonant optical conditions, the amplitude of~the inverted peak may either increase or~decrease. This relation depends on~the value of~$I$.

When a~noticeable part of~the atomic population is~concentrated at~the non-absorbing magnetic sublevels, the two-photon detuning leads to~a~smaller increase in~the background absorption than at~the optical resonance, thereby effectively enhancing its amplitude. Similar to~the crossover case, the resonance becomes narrower as~its amplitude increases, which is~attributed to~smaller optical pumping of~the non-absorbing sublevels. These effects are most pronounced for $I = 3/2$, where $^{87}$Rb is~one of~the cases. The optical wave $\vec{\mathcal{E}}_1$ does not induce transitions from sublevels forming $\Lambda$-schemes by~$\vec{\mathcal{E}}_2$ (the end ones and $F_g=1,\,m_{F_g}=0$), so~they are unaffected in~contrast to~atoms with other~$I$. Therefore, this system of~transitions allows to~trap the most possible amount of~atoms on~these sublevels. All other ground-state sublevels will be~unpopulated, if~we~consider the steady-state regime. This special case will not occur for other values of~$I$.

Considering absolute value of~the Rabi frequencies product for hyperfine $\Lambda$-schemes between sublevels $F^\downarrow_g,\,m_F = |I - 3/2|$ and $F^\uparrow_g,\,m_F = |I + 1/2|$, its value is~practically the same for $I\in[1,\,4]$ undergoing a~slow decay after the maximum at~$I=2$. But, Rabi frequencies from sublevels $F^\downarrow_g,\,m_F = |I - 3/2|$ grow with nuclear spin value $\propto\left[(3-2I)/(1+2I)\right]/\sqrt{3}$, which means that $\Lambda$-schemes become more affected and the less atoms are trapped in the corresponding superpositions. Also, there are more sublevels for larger $I$. This implies that for the same optical pumping rate a~greater flight time is~required for the waves to~optically pump atoms to~the end sublevels and corresponding $\Lambda$-schemes formed by~$\sigma$ transitions. It~can be~concluded that with increasing $I$ the increment of~the amplitude of~the low-frequency eigen peak caused by~the two-photon detuning becomes weaker. After a~certain value of $I$, the absorption in~the center of~the resonance peak increases less compared to~the Doppler background, which leads to~a~decrease in~the amplitude under two-photon detuning.

\begin{figure}[t] 
  \centering
\includegraphics[width=1\columnwidth]{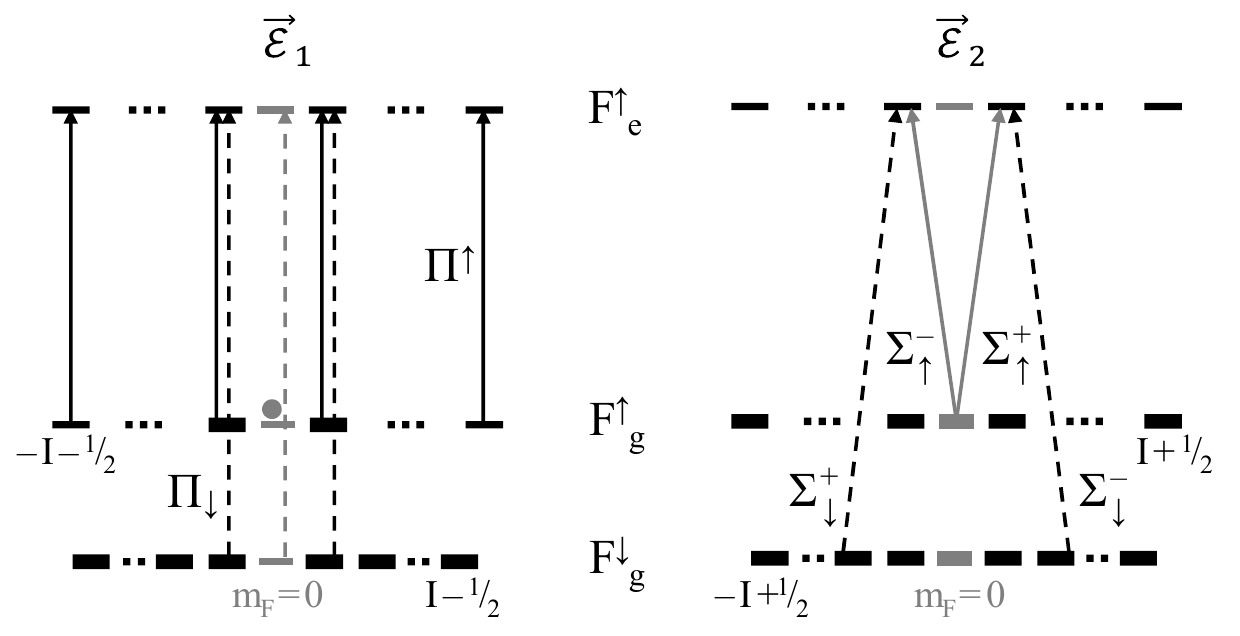}
  \caption{Schemes of~energy levels with $\pi$ (left) and $\sigma$ (right) transitions to~$F^{\uparrow}_e$ state induced by~the optical fields in~lin$\,\perp\,$lin configuration. For clarity, only some of~the allowed transitions are shown.}
  \label{Highfreq}
\end{figure}

\subsection{The high-frequency eigen peak}
\label{High-Frequency-Section}

The analysis of~the eigen peak engaging transitions to~the $F^{\uparrow}_e$ is~similar to~the previous case since optical transitions occur due to~the same selection rules. Under non-resonant optical conditions the main difference lies in~the absence of~non-absorbing states at~the end magnetic sublevels of~$F^\uparrow_g$. The only non-absorbing sublevel in~this case is~$F^\uparrow_g,\,m_F = 0$, which exists only in bosons; see Fig.~\ref{Highfreq}.

Identity~\eqref{PiAndSigmaProducts} also holds for transitions to~$F^{\uparrow}_e$, therefore, at~the optical resonance, the absorption from the sublevels with $m_{F^\uparrow_g} = m_{F^\downarrow_g}$ grows---the corresponding dark superposition of~states induced by~the optical fields are out of~phase.
In~fermions, due to~the absence of~non-absorbing sublevel, the two-photon detuning leads to~a~reduction in~the amplitude of the peak, as~the absorption increases only in~the Doppler background. In~bosons, at~the exact optical resonance the two-photon detuning increases absorption for $I\geq5/2$. This is~due to~the fact that the field $\vec{\mathcal{E}}_2$ forms $\Lambda$-schemes through~the lower level of~the ground state with $m_{F_g}=\pm2$, which do~not exist for smaller value of~the nuclear spin. However, the change in~the absorption at~the exact optical resonance is~smaller compared to~the low-frequency peak and it~diminishes with $I$ due to~lesser population of~the single non-absorbing sublevel. Therefore, two-photon detuning causes only the decrease in~the amplitude.

\section{Experiment}

\begin{figure}[b]
  \centering
\includegraphics[width=1\columnwidth]{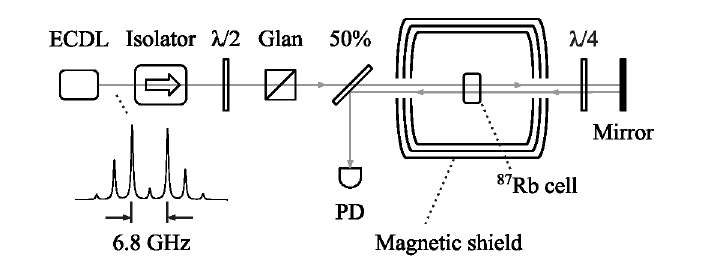}
  \caption{Scheme of~the experimental setup. \hbox{ECDL---extended cavity diode laser}, $\lambda/2$---half-wave plate, \hbox{$\lambda/4$---quarter-wave plate}, PD---photodiode. The inset displays ECDL spectrum and sidebands used for~$^{87}$Rb spectroscopy.}
  \label{ExpSetup}
\end{figure}

The experimental setup~is presented in~Fig.~\ref{ExpSetup}. 
Two extended-cavity diode lasers (ECDL) emitting at~$795$~nm (Rb~D$_1$ line) and $895$~nm (Cs~D$_1$ line) were employed.
Lasers incorporated a~selective element enabling coarse wavelength tuning. 
For experiments involving~$^{87}$Rb and~$^{85}$Rb atoms, an~interference filter with a~bandwidth of~approximately~$100$~GHz was used, and optical feedback was provided by~the cat’s eye configuration utilizing an~output mirror. 
For experiments involving Cs, the ECDL was assembled with a~diffraction grating in~the Littrow scheme.
The output mirror (diffraction grating) was mounted on~a~piezoelectric transducer allowing precise frequency tuning.

\begin{figure*}[t] 
\centering
\includegraphics[width=1\textwidth]{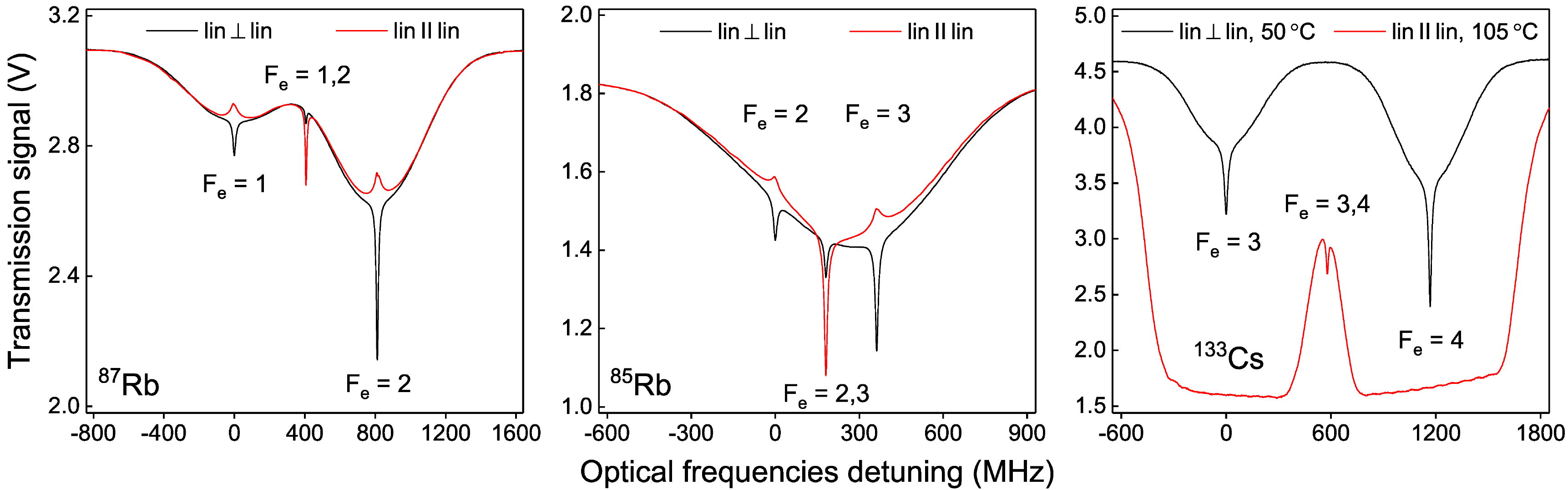}
  \caption{Dual-frequency Doppler-free spectra of~$^{87}$Rb,~$^{85}$Rb and Cs D$_1$ line at~orthogonal (black lines) and parallel (red lines) polarizations of~the counter-propagating optical waves. The horizontal axes represent detuning of~the resonant sidebands from transitions to $F^\downarrow_e$. The laser light intensity is~$2$~mW/cm$^2$ for $^{87,\,85}$Rb and~$16$~mW/cm$^2$ for Cs. The two-photon detuning is~zero.}
  \label{Spectra}
\end{figure*}

\begin{figure*}[t]
  \center
\includegraphics[width=1\textwidth]{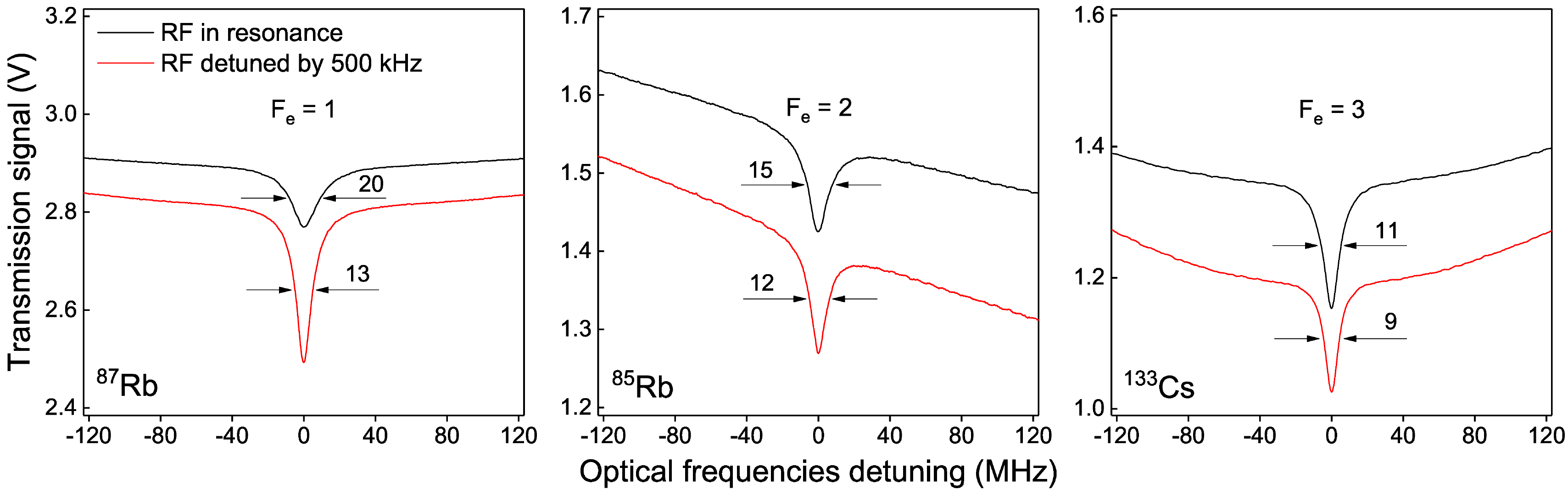}
   \caption{Dual-frequency Doppler-free spectra of~$F^{\downarrow}_e$ eigen peaks taken when microwave frequency was in~resonance with the hyperfine splitting (black lines) and detuned by~$500$~kHz (red lines). FWHM values (in~MHz) are given near to~the resonances. The laser light intensity is~$2$~mW/cm$^2$, the atomic cell temperature is~$50$~$^\circ$C.}
  \label{LFs}
\end{figure*}

The dual-frequency optical field was produced by~microwave modulation of~the ECDL injection current. 
The modulation frequency was set close to~half of~the ground-state hyperfine splitting for~Cs ($4.596$~GHz) and~$^{87}$Rb ($3.417$~GHz), and close~to the full ground-state splitting in~$^{85}$Rb ($3.035$~GHz).
Consequently, the first-order sidebands were resonant with the corresponding D$_1$ line transitions in~Cs and~$^{87}$Rb, whereas the carrier and one of~the first-order sidebands were employed~in experiments with~$^{85}$Rb. All results reported below were obtained with over~$60\%$ of the laser power concentrated in~the resonant sidebands and the ratio of~their amplitudes was close to~$1$. This sideband-to-carrier power ratio was achieved by~matching the external cavity's longitudinal mode spacing with the modulation frequency by~adjustment of~the resonator length~\cite{zibrov2020modulation}.

After passing the optical isolator, the wave went through a~$\lambda/2$ plate and a~polarizer both used to~control the optical power. Then it~was directed into the atomic cell, reflected back by~a~mirror, and registered by~a~photodetector. Polarizations of~the counter-propagating waves were made either mutually orthogonal or~parallel by~rotating a~$\lambda/4$ plate positioned after the cell. The cylindrical atomic cell, filled with alkali metal vapor and equipped with a~heater, was placed inside a~three-layer $\mu$-metal shield to~suppress the external magnetic field. The cell temperature was maintained at~the desired setpoint within the range of~$50–110$~$^\circ$C, with a~precision of~$\pm0.01$~$^\circ$C. The internal lengths of~the cells containing~$^{87}$Rb and~$^{85}$Rb vapor were~$8$~mm each, whereas the cell containing Cs vapor was~$3$~mm long. These lengths were deliberately chosen to~be~shorter than half of~the wavelength of~the microwave transitions between hyperfine ground states: $22$~mm for~$^{87}$Rb, $50$~mm for~$^{85}$Rb, and $16$~mm for Cs. This length allowed to~maximize the effect of~the two-photon detuning by~choosing the proper placing of~the cell along the optical axis~\cite{DFDF87Rb}. All experimental results were obtained with the position yielding the maximum resonance amplitude.

Fig.~\ref{Spectra} illustrates the absorption spectra for~$^{87}$Rb, $^{85}$Rb, and Cs, displaying photodetected signals as~functions of~the synchronous detuning of~the first-order sidebands from transitions to~$F^\downarrow_e$. Consistent with previous studies, all eigen peaks are inverted, exhibiting widths ranging from approximately~$12$ to~$20$~MHz. Spectra for both rubidium isotopes were recorded at~the cell temperature of~$50$~$^\circ$C, sufficient for observing the crossover. They demonstrate that switching from the lin$\,\perp\,$lin configuration to lin$\,||\,$lin results in~an~increased amplitude of~the crossover, while the eigen ones transform to~conventional Doppler-free transmission peaks with noticeably broader widths and smaller amplitudes. It~is~worth noting that the amplitude of~the crossover~in~$^{85}$Rb exceeds even the high-frequency eigen peak. This is~due to~the small hyperfine splitting of~the excited state. Due to~a~significantly larger ratio of~this
splitting to~the Doppler width in~Cs, registration of~the crossover required higher alkali metal concentration. Therefore, the cell was heated up~to~$105$~$^\circ$C. To~achieve a~clearly detectable signal for~the crossover at~this temperature, the laser intensity was increased to~$16$~mW/cm$^2$.

Under these conditions, the measured widths of~the crossovers were approximately~$10$~MHz for~$^{87}$Rb and~$14$~MHz for both~$^{85}$Rb and Cs. 
These values are about twice greater than the natural widths of~the corresponding atomic transitions.
Furthermore, these crossovers are approximately~$1.5$ to~$2$ times narrower than the eigen peaks observed in~the lin$\,\perp\,$lin scheme and significantly narrower than the transmission peaks in~the lin$\,||\,$lin configuration.
These observations confirm the absence of~optical pumping in~the lin$\,||\,$lin scheme for the crossover.

As~discussed in~Section~\ref{Low-Frequency-Section}, eigen peaks with~\hbox{$F^{\downarrow}_e$} and \hbox{$F^{\uparrow}_e$} differ in~the amount of~non-absorbing sublevels for the field $\vec{\mathcal{E}}_1$. The field $\vec{\mathcal{E}}_2$ depopulates these sublevels when the two-photon detuning is~introduced, and the effect differs between the eigen peaks. Also, it~is~important to~evaluate the impact of~the detuning on~the absorption not only at~the exact optical resonance but also at~the Doppler background. Fig.~\ref{LFs} illustrates how the two-photon detuning of~$500$~kHz, sufficient to~eliminate ground-state hyperfine coherences, affects the low-frequency eigen peak in~atoms with different nuclear spin value (the spectra in~the figure are arranged by~increasing the nuclear spin from the left to~right). The measurements were conducted under identical conditions (atomic cell temperature, the laser light intensity) using the lin$\,\perp\,$lin polarization scheme. It~can be~seen from Fig.~\ref{LFs} that for~$^{87}$Rb ($I=3/2$), the resonant absorption increases considerably greater than the Doppler background, resulting in~a~twofold enhancement of~the amplitude. For~$^{85}$Rb ($I=5/2$) the resonant absorption still surpasses the background, although their growths are nearly equivalent. For Cs (\hbox{$I=7/2$}) the change in~the resonant absorption no~longer exceeds that of~the Doppler background, resulting in~a~reduced amplitude. Widths of~all the observed peaks decrease when the two-photon detuning is~applied. Again, the most prominent effect of~the narrowing presents in~$^{87}$Rb as~one can estimate from numbers shown near to~peaks in~Fig.~\ref{LFs}. This comparison confirms that the level structure for the special case $I=3/2$, which enables two coherent superpositions for the field $\vec{\mathcal{E}}_2$ via~$m_F=0$ and~$m_F=\pm2$, provides the most significant optical pumping of~these sublevels.

\section{Summary}

In this paper, we~presented a~generalized analysis of~the physical mechanisms underlying high-contrast Doppler-free absorption resonance related to~system of~transitions \hbox{$J_g=1/2\rightarrow J_e=1/2$} induced by~bichromatic optical fields. It~was carried out in~the basis where the quantization axis is~aligned with the polarization vector of~one of~the counter-propagating waves $\vec{\mathcal{E}}_1$. The analysis is~applicable both for fermions and bosons. The dual-frequency Doppler-free spectra were experimentally investigated in~$^{87}$Rb, $^{85}$Rb, and Cs, which have nuclear spin values of~$I = 3/2, 5/2$ and~$7/2$, respectively.

It was demonstrated that the amplitude of~the crossover resonance significantly increases when the polarizations of the counter-propagating waves are parallel, as~compared to~the orthogonal ones. This enhancement stems from the involvement of~all ground-state Zeeman sublevels in~the absorption and their uniform repopulation. The absence of~optical pumping, one of~the primary mechanisms of~broadening, leads to~a~narrow width of~the crossover. The experimentally measured widths were found to~be~close to~the natural linewidths of~corresponding atoms. These effects are nuclear-spin-independent. 

For eigen peaks, we~analyzed the effect of~the two-photon detuning on~the absorption at~the exact optical resonance and the Doppler background. Transitions to~the excited-state level with total angular momentum \hbox{$F_e=I-1/2$} have at~least two dark sublevels which accumulate atoms for the field $\vec{\mathcal{E}}_1$. It~was explained that the counter-propagating wave induces dark-state superpositions at~these sublevels. The amount of~atoms, which can be~accumulated at~these sublevles, decreases with $I$. Energy configuration of~atoms with~$I = 3/2$ possesses $3$ of~such sublevels and $\Lambda$-schemes linking them are unaffected by~the wave $\vec{\mathcal{E}}_1$. Hence, the two-photon detuning changes the resonant absorption stronger than the background one. The comparison of~experimental results revealed that the two-photon detuning doubled the amplitude value of~this resonance in~$^{87}$Rb. Only slight amplitude growth was observed in~$^{85}$Rb, and the amplitude in~Cs decreased because change in~the background absorption exceeded that at~the exact optical resonance.

Considering the transitions to~$F_e=I+1/2$ induced by~the field $\vec{\mathcal{E}}_1$, they give only one non-absorbing sublevel in~bosons or~absence of~them in~fermions. Therefore, the two-photon detuning destroys coherent dark superpositions for the background. At~the exact optical resonance, the change in~the absorption takes place only for $I\geq5/2$. This causes mostly growth of~the background absorption tremendously exceeding the resonant one. Thus, the amplitude of~the high-frequency eigen peak decreased for all the atoms under investigation to~a~comparable extent. 

\section*{Funding Information}

The authors receive funding from Russian Science Foundation (grant No. 24-72-10134).


\begin{thebibliography}{16}%
\makeatletter
\providecommand \@ifxundefined [1]{%
 \@ifx{#1\undefined}
}%
\providecommand \@ifnum [1]{%
 \ifnum #1\expandafter \@firstoftwo
 \else \expandafter \@secondoftwo
 \fi
}%
\providecommand \@ifx [1]{%
 \ifx #1\expandafter \@firstoftwo
 \else \expandafter \@secondoftwo
 \fi
}%
\providecommand \natexlab [1]{#1}%
\providecommand \enquote  [1]{``#1''}%
\providecommand \bibnamefont  [1]{#1}%
\providecommand \bibfnamefont [1]{#1}%
\providecommand \citenamefont [1]{#1}%
\providecommand \href@noop [0]{\@secondoftwo}%
\providecommand \href [0]{\begingroup \@sanitize@url \@href}%
\providecommand \@href[1]{\@@startlink{#1}\@@href}%
\providecommand \@@href[1]{\endgroup#1\@@endlink}%
\providecommand \@sanitize@url [0]{\catcode `\\12\catcode `\$12\catcode `\&12\catcode `\#12\catcode `\^12\catcode `\_12\catcode `\%12\relax}%
\providecommand \@@startlink[1]{}%
\providecommand \@@endlink[0]{}%
\providecommand \url  [0]{\begingroup\@sanitize@url \@url }%
\providecommand \@url [1]{\endgroup\@href {#1}{\urlprefix }}%
\providecommand \urlprefix  [0]{URL }%
\providecommand \Eprint [0]{\href }%
\providecommand \doibase [0]{http://dx.doi.org/}%
\providecommand \selectlanguage [0]{\@gobble}%
\providecommand \bibinfo  [0]{\@secondoftwo}%
\providecommand \bibfield  [0]{\@secondoftwo}%
\providecommand \translation [1]{[#1]}%
\providecommand \BibitemOpen [0]{}%
\providecommand \bibitemStop [0]{}%
\providecommand \bibitemNoStop [0]{.\EOS\space}%
\providecommand \EOS [0]{\spacefactor3000\relax}%
\providecommand \BibitemShut  [1]{\csname bibitem#1\endcsname}%
\let\auto@bib@innerbib\@empty
\bibitem [{\citenamefont {Letokhov}\ and\ \citenamefont {Chebotaev}(1977)}]{letokhov1977nonlinear}%
  \BibitemOpen
  \bibfield  {author} {\bibinfo {author} {\bibfnamefont {V.}~\bibnamefont {Letokhov}}\ and\ \bibinfo {author} {\bibfnamefont {V.}~\bibnamefont {Chebotaev}},\ }\href {https://books.google.ru/books?id=lcfvAAAAMAAJ} {\emph {\bibinfo {title} {Nonlinear Laser Spectroscopy}}},\ Springer series in optical sciences\ (\bibinfo  {publisher} {Springer-Verlag},\ \bibinfo {year} {1977})\BibitemShut {NoStop}%
\bibitem [{\citenamefont {Schawlow}(1982)}]{schawlow1982spectroscopy}%
  \BibitemOpen
  \bibfield  {author} {\bibinfo {author} {\bibfnamefont {A.~L.}\ \bibnamefont {Schawlow}},\ }\href@noop {} {\bibfield  {journal} {\bibinfo  {journal} {Reviews of Modern Physics}\ }\textbf {\bibinfo {volume} {54}},\ \bibinfo {pages} {697} (\bibinfo {year} {1982})}\BibitemShut {NoStop}%
\bibitem [{\citenamefont {H{\"a}nsch}(1977)}]{hansch1977high}%
  \BibitemOpen
  \bibfield  {author} {\bibinfo {author} {\bibfnamefont {T.~W.}\ \bibnamefont {H{\"a}nsch}},\ }\href@noop {} {\bibfield  {journal} {\bibinfo  {journal} {Physics Today}\ }\textbf {\bibinfo {volume} {30}},\ \bibinfo {pages} {34} (\bibinfo {year} {1977})}\BibitemShut {NoStop}%
\bibitem [{\citenamefont {Newman}\ \emph {et~al.}(2019)\citenamefont {Newman}, \citenamefont {Maurice}, \citenamefont {Drake}, \citenamefont {Stone}, \citenamefont {Briles}, \citenamefont {Spencer}, \citenamefont {Fredrick}, \citenamefont {Li}, \citenamefont {Westly}, \citenamefont {Ilic} \emph {et~al.}}]{newman2019architecture}%
  \BibitemOpen
  \bibfield  {author} {\bibinfo {author} {\bibfnamefont {Z.~L.}\ \bibnamefont {Newman}}, \bibinfo {author} {\bibfnamefont {V.}~\bibnamefont {Maurice}}, \bibinfo {author} {\bibfnamefont {T.}~\bibnamefont {Drake}}, \bibinfo {author} {\bibfnamefont {J.~R.}\ \bibnamefont {Stone}}, \bibinfo {author} {\bibfnamefont {T.~C.}\ \bibnamefont {Briles}}, \bibinfo {author} {\bibfnamefont {D.~T.}\ \bibnamefont {Spencer}}, \bibinfo {author} {\bibfnamefont {C.}~\bibnamefont {Fredrick}}, \bibinfo {author} {\bibfnamefont {Q.}~\bibnamefont {Li}}, \bibinfo {author} {\bibfnamefont {D.}~\bibnamefont {Westly}}, \bibinfo {author} {\bibfnamefont {B.~R.}\ \bibnamefont {Ilic}},  \emph {et~al.},\ }\href@noop {} {\bibfield  {journal} {\bibinfo  {journal} {Optica}\ }\textbf {\bibinfo {volume} {6}},\ \bibinfo {pages} {680} (\bibinfo {year} {2019})}\BibitemShut {NoStop}%
\bibitem [{\citenamefont {Martin}\ \emph {et~al.}(2018)\citenamefont {Martin}, \citenamefont {Phelps}, \citenamefont {Lemke}, \citenamefont {Bigelow}, \citenamefont {Stuhl}, \citenamefont {Wojcik}, \citenamefont {Holt}, \citenamefont {Coddington}, \citenamefont {Bishop},\ and\ \citenamefont {Burke}}]{PhysRevApplied.9.014019}%
  \BibitemOpen
  \bibfield  {author} {\bibinfo {author} {\bibfnamefont {K.~W.}\ \bibnamefont {Martin}}, \bibinfo {author} {\bibfnamefont {G.}~\bibnamefont {Phelps}}, \bibinfo {author} {\bibfnamefont {N.~D.}\ \bibnamefont {Lemke}}, \bibinfo {author} {\bibfnamefont {M.~S.}\ \bibnamefont {Bigelow}}, \bibinfo {author} {\bibfnamefont {B.}~\bibnamefont {Stuhl}}, \bibinfo {author} {\bibfnamefont {M.}~\bibnamefont {Wojcik}}, \bibinfo {author} {\bibfnamefont {M.}~\bibnamefont {Holt}}, \bibinfo {author} {\bibfnamefont {I.}~\bibnamefont {Coddington}}, \bibinfo {author} {\bibfnamefont {M.~W.}\ \bibnamefont {Bishop}}, \ and\ \bibinfo {author} {\bibfnamefont {J.~H.}\ \bibnamefont {Burke}},\ }\href {\doibase 10.1103/PhysRevApplied.9.014019} {\bibfield  {journal} {\bibinfo  {journal} {Phys. Rev. Appl.}\ }\textbf {\bibinfo {volume} {9}},\ \bibinfo {pages} {014019} (\bibinfo {year} {2018})}\BibitemShut {NoStop}%
\bibitem [{\citenamefont {Perrella}\ \emph {et~al.}(2019)\citenamefont {Perrella}, \citenamefont {Light}, \citenamefont {Anstie}, \citenamefont {Baynes}, \citenamefont {White},\ and\ \citenamefont {Luiten}}]{PhysRevApplied.12.054063}%
  \BibitemOpen
  \bibfield  {author} {\bibinfo {author} {\bibfnamefont {C.}~\bibnamefont {Perrella}}, \bibinfo {author} {\bibfnamefont {P.}~\bibnamefont {Light}}, \bibinfo {author} {\bibfnamefont {J.}~\bibnamefont {Anstie}}, \bibinfo {author} {\bibfnamefont {F.}~\bibnamefont {Baynes}}, \bibinfo {author} {\bibfnamefont {R.}~\bibnamefont {White}}, \ and\ \bibinfo {author} {\bibfnamefont {A.}~\bibnamefont {Luiten}},\ }\href {\doibase 10.1103/PhysRevApplied.12.054063} {\bibfield  {journal} {\bibinfo  {journal} {Phys. Rev. Appl.}\ }\textbf {\bibinfo {volume} {12}},\ \bibinfo {pages} {054063} (\bibinfo {year} {2019})}\BibitemShut {NoStop}%
\bibitem [{\citenamefont {Beard}\ \emph {et~al.}(2024)\citenamefont {Beard}, \citenamefont {Martin}, \citenamefont {Elgin}, \citenamefont {Kasch},\ and\ \citenamefont {Krzyzewski}}]{Beard:24}%
  \BibitemOpen
  \bibfield  {author} {\bibinfo {author} {\bibfnamefont {R.}~\bibnamefont {Beard}}, \bibinfo {author} {\bibfnamefont {K.~W.}\ \bibnamefont {Martin}}, \bibinfo {author} {\bibfnamefont {J.~D.}\ \bibnamefont {Elgin}}, \bibinfo {author} {\bibfnamefont {B.~L.}\ \bibnamefont {Kasch}}, \ and\ \bibinfo {author} {\bibfnamefont {S.~P.}\ \bibnamefont {Krzyzewski}},\ }\href {\doibase 10.1364/OE.513974} {\bibfield  {journal} {\bibinfo  {journal} {Opt. Express}\ }\textbf {\bibinfo {volume} {32}},\ \bibinfo {pages} {7417} (\bibinfo {year} {2024})}\BibitemShut {NoStop}%
\bibitem [{\citenamefont {Ruelle}\ \emph {et~al.}(2024)\citenamefont {Ruelle}, \citenamefont {Batori}, \citenamefont {Kundermann}, \citenamefont {Stehlin}, \citenamefont {Helson}, \citenamefont {Haesler}, \citenamefont {Lecomte}, \citenamefont {Droz},\ and\ \citenamefont {Karlen}}]{10722354}%
  \BibitemOpen
  \bibfield  {author} {\bibinfo {author} {\bibfnamefont {T.}~\bibnamefont {Ruelle}}, \bibinfo {author} {\bibfnamefont {E.}~\bibnamefont {Batori}}, \bibinfo {author} {\bibfnamefont {S.}~\bibnamefont {Kundermann}}, \bibinfo {author} {\bibfnamefont {X.}~\bibnamefont {Stehlin}}, \bibinfo {author} {\bibfnamefont {V.}~\bibnamefont {Helson}}, \bibinfo {author} {\bibfnamefont {J.}~\bibnamefont {Haesler}}, \bibinfo {author} {\bibfnamefont {S.}~\bibnamefont {Lecomte}}, \bibinfo {author} {\bibfnamefont {F.}~\bibnamefont {Droz}}, \ and\ \bibinfo {author} {\bibfnamefont {S.}~\bibnamefont {Karlen}},\ }in\ \href {\doibase 10.1109/EFTF61992.2024.10722354} {\emph {\bibinfo {booktitle} {2024 European Frequency and Time Forum (EFTF)}}}\ (\bibinfo {year} {2024})\ pp.\ \bibinfo {pages} {1--2}\BibitemShut {NoStop}%
\bibitem [{\citenamefont {Duspayev}\ \emph {et~al.}(2024)\citenamefont {Duspayev}, \citenamefont {Owens}, \citenamefont {Dash},\ and\ \citenamefont {Raithel}}]{Duspayev_2024}%
  \BibitemOpen
  \bibfield  {author} {\bibinfo {author} {\bibfnamefont {A.}~\bibnamefont {Duspayev}}, \bibinfo {author} {\bibfnamefont {C.}~\bibnamefont {Owens}}, \bibinfo {author} {\bibfnamefont {B.}~\bibnamefont {Dash}}, \ and\ \bibinfo {author} {\bibfnamefont {G.}~\bibnamefont {Raithel}},\ }\href {\doibase 10.1088/2058-9565/ad77ef} {\bibfield  {journal} {\bibinfo  {journal} {Quantum Science and Technology}\ }\textbf {\bibinfo {volume} {9}},\ \bibinfo {pages} {045046} (\bibinfo {year} {2024})}\BibitemShut {NoStop}%
\bibitem [{\citenamefont {Hafiz}\ \emph {et~al.}(2016)\citenamefont {Hafiz}, \citenamefont {Coget}, \citenamefont {De~Clercq},\ and\ \citenamefont {Boudot}}]{hafiz2016doppler}%
  \BibitemOpen
  \bibfield  {author} {\bibinfo {author} {\bibfnamefont {M.~A.}\ \bibnamefont {Hafiz}}, \bibinfo {author} {\bibfnamefont {G.}~\bibnamefont {Coget}}, \bibinfo {author} {\bibfnamefont {E.}~\bibnamefont {De~Clercq}}, \ and\ \bibinfo {author} {\bibfnamefont {R.}~\bibnamefont {Boudot}},\ }\href@noop {} {\bibfield  {journal} {\bibinfo  {journal} {Optics letters}\ }\textbf {\bibinfo {volume} {41}},\ \bibinfo {pages} {2982} (\bibinfo {year} {2016})}\BibitemShut {NoStop}%
\bibitem [{\citenamefont {Hafiz}\ \emph {et~al.}(2017)\citenamefont {Hafiz}, \citenamefont {Brazhnikov}, \citenamefont {Coget}, \citenamefont {Taichenachev}, \citenamefont {Yudin}, \citenamefont {De~Clercq},\ and\ \citenamefont {Boudot}}]{hafiz2017high}%
  \BibitemOpen
  \bibfield  {author} {\bibinfo {author} {\bibfnamefont {M.~A.}\ \bibnamefont {Hafiz}}, \bibinfo {author} {\bibfnamefont {D.}~\bibnamefont {Brazhnikov}}, \bibinfo {author} {\bibfnamefont {G.}~\bibnamefont {Coget}}, \bibinfo {author} {\bibfnamefont {A.}~\bibnamefont {Taichenachev}}, \bibinfo {author} {\bibfnamefont {V.}~\bibnamefont {Yudin}}, \bibinfo {author} {\bibfnamefont {E.}~\bibnamefont {De~Clercq}}, \ and\ \bibinfo {author} {\bibfnamefont {R.}~\bibnamefont {Boudot}},\ }\href@noop {} {\bibfield  {journal} {\bibinfo  {journal} {New Journal of Physics}\ }\textbf {\bibinfo {volume} {19}},\ \bibinfo {pages} {073028} (\bibinfo {year} {2017})}\BibitemShut {NoStop}%
\bibitem [{\citenamefont {Zhao}\ \emph {et~al.}(2021)\citenamefont {Zhao}, \citenamefont {Jiang}, \citenamefont {Fang}, \citenamefont {Qiu}, \citenamefont {Ma}, \citenamefont {Han}, \citenamefont {Lu},\ and\ \citenamefont {Lee}}]{zhao2021laser}%
  \BibitemOpen
  \bibfield  {author} {\bibinfo {author} {\bibfnamefont {M.}~\bibnamefont {Zhao}}, \bibinfo {author} {\bibfnamefont {X.}~\bibnamefont {Jiang}}, \bibinfo {author} {\bibfnamefont {R.}~\bibnamefont {Fang}}, \bibinfo {author} {\bibfnamefont {Y.}~\bibnamefont {Qiu}}, \bibinfo {author} {\bibfnamefont {Z.}~\bibnamefont {Ma}}, \bibinfo {author} {\bibfnamefont {C.}~\bibnamefont {Han}}, \bibinfo {author} {\bibfnamefont {B.}~\bibnamefont {Lu}}, \ and\ \bibinfo {author} {\bibfnamefont {C.}~\bibnamefont {Lee}},\ }\href@noop {} {\bibfield  {journal} {\bibinfo  {journal} {Applied Optics}\ }\textbf {\bibinfo {volume} {60}},\ \bibinfo {pages} {5203} (\bibinfo {year} {2021})}\BibitemShut {NoStop}%
\bibitem [{\citenamefont {Gusching}\ \emph {et~al.}(2021)\citenamefont {Gusching}, \citenamefont {Petersen}, \citenamefont {Passilly}, \citenamefont {Brazhnikov}, \citenamefont {Hafiz},\ and\ \citenamefont {Boudot}}]{gusching2021short}%
  \BibitemOpen
  \bibfield  {author} {\bibinfo {author} {\bibfnamefont {A.}~\bibnamefont {Gusching}}, \bibinfo {author} {\bibfnamefont {M.}~\bibnamefont {Petersen}}, \bibinfo {author} {\bibfnamefont {N.}~\bibnamefont {Passilly}}, \bibinfo {author} {\bibfnamefont {D.}~\bibnamefont {Brazhnikov}}, \bibinfo {author} {\bibfnamefont {M.~A.}\ \bibnamefont {Hafiz}}, \ and\ \bibinfo {author} {\bibfnamefont {R.}~\bibnamefont {Boudot}},\ }\href@noop {} {\bibfield  {journal} {\bibinfo  {journal} {JOSA B}\ }\textbf {\bibinfo {volume} {38}},\ \bibinfo {pages} {3254} (\bibinfo {year} {2021})}\BibitemShut {NoStop}%
\bibitem [{\citenamefont {Gusching}\ \emph {et~al.}(2023)\citenamefont {Gusching}, \citenamefont {Petersen}, \citenamefont {Passilly}, \citenamefont {Brazhnikov}, \citenamefont {Hafiz},\ and\ \citenamefont {Boudot}}]{gusching2023short}%
  \BibitemOpen
  \bibfield  {author} {\bibinfo {author} {\bibfnamefont {A.}~\bibnamefont {Gusching}}, \bibinfo {author} {\bibfnamefont {M.}~\bibnamefont {Petersen}}, \bibinfo {author} {\bibfnamefont {N.}~\bibnamefont {Passilly}}, \bibinfo {author} {\bibfnamefont {D.}~\bibnamefont {Brazhnikov}}, \bibinfo {author} {\bibfnamefont {M.~A.}\ \bibnamefont {Hafiz}}, \ and\ \bibinfo {author} {\bibfnamefont {R.}~\bibnamefont {Boudot}},\ }\href@noop {} {\bibfield  {journal} {\bibinfo  {journal} {JOSA B}\ }\textbf {\bibinfo {volume} {40}},\ \bibinfo {pages} {501} (\bibinfo {year} {2023})}\BibitemShut {NoStop}%
\bibitem [{\citenamefont {Velichansky}\ \emph {et~al.}(2025)\citenamefont {Velichansky}, \citenamefont {Sabakar}, \citenamefont {Fedorov}, \citenamefont {Tsygankov}, \citenamefont {Chuchelov}, \citenamefont {Vaskovskaya}, \citenamefont {Vassiliev},\ and\ \citenamefont {Zibrov}}]{DFDF87Rb}%
  \BibitemOpen
  \bibfield  {author} {\bibinfo {author} {\bibfnamefont {V.~L.}\ \bibnamefont {Velichansky}}, \bibinfo {author} {\bibfnamefont {K.~M.}\ \bibnamefont {Sabakar}}, \bibinfo {author} {\bibfnamefont {A.~S.}\ \bibnamefont {Fedorov}}, \bibinfo {author} {\bibfnamefont {E.~A.}\ \bibnamefont {Tsygankov}}, \bibinfo {author} {\bibfnamefont {D.~S.}\ \bibnamefont {Chuchelov}}, \bibinfo {author} {\bibfnamefont {M.~I.}\ \bibnamefont {Vaskovskaya}}, \bibinfo {author} {\bibfnamefont {V.~V.}\ \bibnamefont {Vassiliev}}, \ and\ \bibinfo {author} {\bibfnamefont {S.~A.}\ \bibnamefont {Zibrov}},\ }\href {\doibase 10.1134/S0021364025605615} {\bibfield  {journal} {\bibinfo  {journal} {JETP Letters}\ }\textbf {\bibinfo {volume} {121}} (\bibinfo {year} {2025}),\ 10.1134/S0021364025605615}\BibitemShut {NoStop}%
\bibitem [{\citenamefont {Zibrov}\ \emph {et~al.}(2020)\citenamefont {Zibrov}, \citenamefont {Chuchelov}, \citenamefont {Drakin}, \citenamefont {Shiryaev}, \citenamefont {Tsygankov}, \citenamefont {Vaskovskaya}, \citenamefont {Vassiliev}, \citenamefont {Velichansky},\ and\ \citenamefont {Bogatov}}]{zibrov2020modulation}%
  \BibitemOpen
  \bibfield  {author} {\bibinfo {author} {\bibfnamefont {S.~A.}\ \bibnamefont {Zibrov}}, \bibinfo {author} {\bibfnamefont {D.~S.}\ \bibnamefont {Chuchelov}}, \bibinfo {author} {\bibfnamefont {A.~E.}\ \bibnamefont {Drakin}}, \bibinfo {author} {\bibfnamefont {D.~A.}\ \bibnamefont {Shiryaev}}, \bibinfo {author} {\bibfnamefont {E.~A.}\ \bibnamefont {Tsygankov}}, \bibinfo {author} {\bibfnamefont {M.~I.}\ \bibnamefont {Vaskovskaya}}, \bibinfo {author} {\bibfnamefont {V.~V.}\ \bibnamefont {Vassiliev}}, \bibinfo {author} {\bibfnamefont {V.~L.}\ \bibnamefont {Velichansky}}, \ and\ \bibinfo {author} {\bibfnamefont {A.~P.}\ \bibnamefont {Bogatov}},\ }\href@noop {} {\bibfield  {journal} {\bibinfo  {journal} {IEEE Journal of Quantum Electronics}\ }\textbf {\bibinfo {volume} {56}},\ \bibinfo {pages} {1} (\bibinfo {year} {2020})}\BibitemShut {NoStop}%
\end{thebibliography}
\end{document}